\documentclass{article}[12pt]

\usepackage{graphicx}

\begin{document}

\title{Another type of log-periodic oscillations on Polish stock market?} 
 
\author{ Piotr Gnaci\'nski\footnote{pg@iftia.univ.gda.pl} and 	Danuta Makowiec\footnote{fizdm@univ.gda.pl} \\
        Institute of Theoretical Physics and Astrophysics,  
  Gda\'nsk University \\ ul.Wita Stwosza 57, 80-952~Gda\'nsk, Poland 
}
 
\maketitle

\begin{abstract}   
 Log-periodic oscillations have been used to predict price trends and
crashes on financial markets. So far two types of log-periodic 
oscillations have been associated with the real markets. 
The first type are oscillations which accompany a rising market 
and which ends in a crash. The second type oscillations, called "anti-bubbles" appear after a crash, when the prices decreases.
Here, we propose the third type of log-periodic oscillations, where a exogenous crash initializes a log-periodic behavior of market, and the market is growing up. Such behavior has been identified on Polish stock market index between the "Russian crisis" (August 1998) and the "New Economy crash" in April 2000.

\end{abstract}


\section{Introduction}

Starting from 1996, Sornette {\it et al.} \cite{First} in the series of papers
\cite{Large,Nasdaq2000,anti-bubbles,Bubbles,Exogenous,predicting,comment,outliers,critical},
give arguments that crashes are analogous to critical points, which are preceded by 
log-periodic oscillations. Such oscillations were studied in statistical 
physics. The interactions between investors lead to a speculative "bubble" which ends in a crash. 
Over 50 such crashes have been described on stock, FX and Gold
market.
Moreover, using the symmetric in time bubble formula they have found 
so called "anti-bubble" price development which describes how a market behaves 
after a crash. Starting at rapidly oscillating state after the crash a market goes
down with its price decorated by log-periodic oscillations.
Both types of oscillations: bubble and anti-bubble are assigned to herding 
behavior of investors.
However difficulties have been encountered when crashes on Eastern
Europe stock markets were studied \cite{Bubbles}.
We propose another type of oscillations, where an exogenous crash 
initializes a log-periodic price development, but the prices are rising.
We will call it "inverted bubble".

In order to accurately define a crash we use so called "drawdowns".
A drawdown is a persistent decrease in the price over consecutive 
days \cite{comment}. We ignore corrections between local maximum 
and local minimum less than $ \epsilon=30\% $ of the preceding price fall.

Fig.1 presents a scheme of a $\epsilon$-drawdown.
In the first step we construct drawdowns (drawups) by merging together
quotations with price changes in the same direction. For example, the first three prices of Fig.1. are merged into a drawdown AB, the next two create a drawup BC, followed by a drawdown CD. In the second step $\epsilon$-drawdowns are constructed as follows.If the drawup BC: $P_C-P_B$, is less than $\epsilon$ (here we assume that $\epsilon=30\%$) of the absolute value of the drawdown AB: $|P_B-P_A|$, and price in the point D is less than in the price in the point B, than the drawup BC is treated as the correction to the trend. So that the  $\epsilon$-drawdown AD is made in place of the two drawdowns AB, and CD and the drawup BC. 
Finally, the price change is transferred into logarithms:
\begin{equation}
drawdown=ln(P_{max}/P_{min}) 
\end{equation}
where $P_{max}$ and $P_{min}$ are prices at beginning and at the end of a drawdown (here points A and D). According to \cite{comment} the distribution of drawdowns is a stretched exponential 
\begin{equation}
f(x)=a \cdot exp(-b|x|^z)
\end{equation}
Fig.2 shows the distribution of drawdowns of WIG (Warsaw Stock Index) 
from 1991 to August 2002. The fit of stretched exponential to this 
distribution  gives us:  $ a=63.6,\quad b=0.019,\quad z=1.08$. Hence , events 
with drawdowns less than -14.5 \% are outliers. We will adopt a definition
of a crash as a drawdown with price loss over 14.5 \%. An extreme rise will be a drawup of more than 17.5 \%.

\section{Log-periodic oscillations}

The prices or index values prior to crash are described by the first term 
"Landau" expansion. The expansion describes a power law behavior
of price $p(\tau)$:
\begin{equation}
\frac{d p(\tau )}{d\ln \tau }=(\beta+i\omega) p(\tau )
\end{equation}
This equation is sometimes written in the form:
\begin{equation}
\frac{d\ln p(\tau )}{d\ln \tau }=(\beta+i\omega)
\end{equation}
which illustrates that the price $p(\tau)$ becomes self-similar with respect 
to the dilation of the distance $\tau$. The relative variations 
$d\ln p=dp/p$ of price with respect to the relative variations of
the time to crash $d\ln \tau=d\tau/\tau$ are independent of time.
From equation 3 one can obtain a log-periodic development of price:
\begin{equation}
p(t)=A+\tau ^\beta \cdot \left[ B+C\cos \left( \omega \ln \tau +\phi \right) \right]
\end{equation}
where $\tau =\left| t_c-t\right| $ , and 
$t_c$ is critical time (the time of crash). 

The numerical way to identify log-periodicity in a time series is to
fit the 7 parameters-function from eq. 5 to the data. The amoeba fitting
procedure \cite{NR} was used to minimize the variance: 
\begin{equation}
Var=\frac{1}{N}\sum_{k=1}^{N}(p_k-p(t_k;A,B,C,\beta,\omega,t_c,\phi))^2
\end{equation}
Where N is the number of points in the data set, $t_k$ and $p_k$ are the time and
price of the k-th point.

To convince that the oscillations are really present 
in the data we transform the analyzed data to a pure cosine 
function \cite{predicting}:
\begin{eqnarray}
t_k\longrightarrow ln|t_k-t_c|  \\
p_k\longrightarrow \frac{p_k-(A+B\tau ^\beta )}{C\tau ^\beta }  
\end{eqnarray}
From the transformed data we made a Lomb periodogram \cite{NR}
The Lomb periodogram is used to find periodicity in unevenly sampled data.
The Lomb periodogram should give us the same frequency as the 
fitting procedure. 

At least four distinct log-oscillation events can be identified on the 
Polish stock market index WIG in the years 1991 to 2002 (Fig. 3): the speculative bubble which ended with the crash in  March 1994, then an anti-bubble recovery from that crash and the inverted bubble crash of April 2000 followed by the anti-bubble price development.

\section{Why inverted log-periodic oscillation?}

The Nasdaq "New Economy" fall in April 2000 was a 
bubble ending in a crash \cite{Nasdaq2000}.
This event is an instance of log-periodic price development on rising
market ending with a crash. A fit with eq. (5)
describes the behavior of Nasdaq from Spring 1997 to April 2000. 
The examined period consists of four evident oscillations.

The "New Economy" crash was also observed on the Polish stock market. 
However no satisfying long-term fit with the eq. (5) could be made. 
Here are our attempts. On Fig.4 the bubble fit to Warsaw Stock Index WIG is shown. The best fit extends over one year, with one oscillation period only, and gives the critical time 4 months after the "New Economy" crash happened. 
On Fig. 5 the fit to the period August 1998 (so called "Russian crisis", 
which caused WIG $drawdown=-31.5\% $) to August 2000 is shown. The critical time is April 2000, but the fit is based on a small correction in January 2000, which itself could not be well approximated. It is impossible to predict this crash with log-periodic function. It was possible to made the presented fit only  because we know when the crash occurred.
Summing up, we claim that both fits are doubtful, and are extended only about 1.5 periods of the  cosine function.

Therefore we propose another possibility of describing the development of WIG index in the time period from  August 1998 to April 2000. On August 1998 the "Russian crisis" occurred. The "Russian crisis" was a exogenous event for the Polish stock market.
The critical time $t_c$ (August 1998) means the starting point of inverted bubble: a log-periodicity price behavior with a {\bf rising} market. Now, the fit extends over 4 periods of the cosine function (Fig. 6). 
We have found two satisfactory fits to the WIG index: 
the first fit is  with $\omega=9.8$ and second one is with $\omega=11.3$. 

Unfortunately, the inverted bubble was not found in Nasdaq data. 
From an inverted bubble one can not predict the time of the next 
crash (April 2000 in this case). However it can be expected, that
crash should coincide with the maximum of the cosine function.

\mbox{} \\
{ \bf Note }

  After finishing this paper we have recognized that inverted
bubbles were independently described by Zhou and Sornette \cite{BullishAntiBubbles}. 
They have found six market indices with such behavior (they call them
"bullish anti-bubbles"). All critical times were set  between August 
and November 2000. However, these log-periodic oscillations have not
ended with a crash yet.

\mbox{} \\
{ \bf Acknowledgments }

  This work was supported by Gda\'nsk University grant BW 5400-5-0014-3.





\begin{figure}
\begin{center}
 \includegraphics[angle=0 , height=5 cm ]{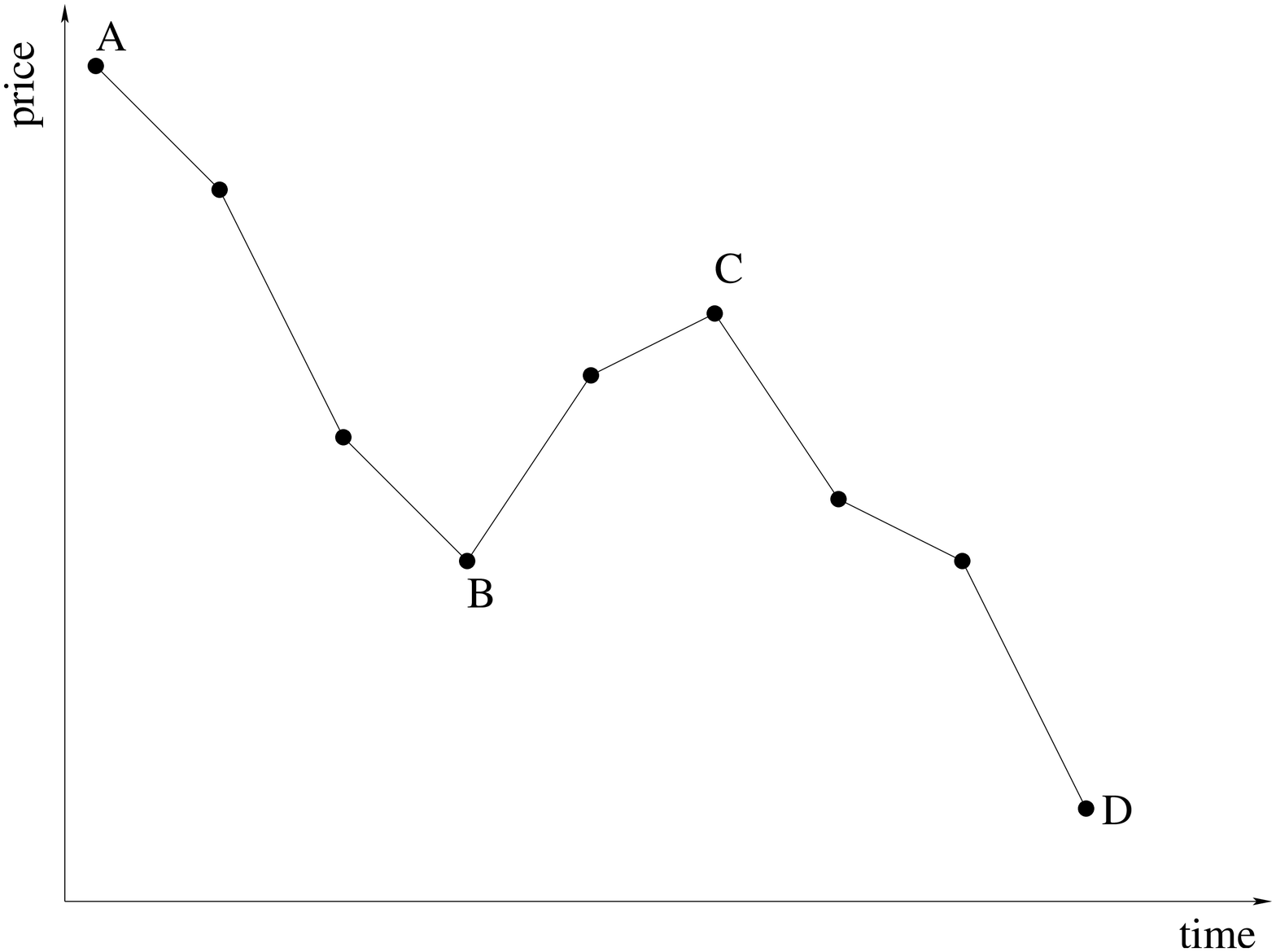}
\end{center}
 \caption{ 
    Construction of $\epsilon$-drawdowns.
	 }
\end{figure}

\begin{figure}
\begin{center}
 \includegraphics[angle=-90 , width=10 cm ]{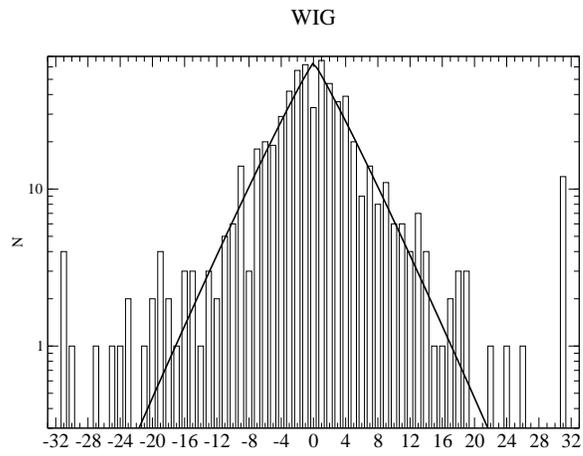}
 \end{center}
 \caption{ 
    The histogram of drawdowns and drawups (in \%) on on Polish stock
    market index WIG from beginning (1991) to August 2002. The solid line 
    is a best fit to the distribution with equation 2:
    $a=63.6$, $b=0.019$, $z=1.08$.
	 }
\end{figure}

\begin{figure}
\begin{center}
 \includegraphics[angle=-90 , width=15 cm ]{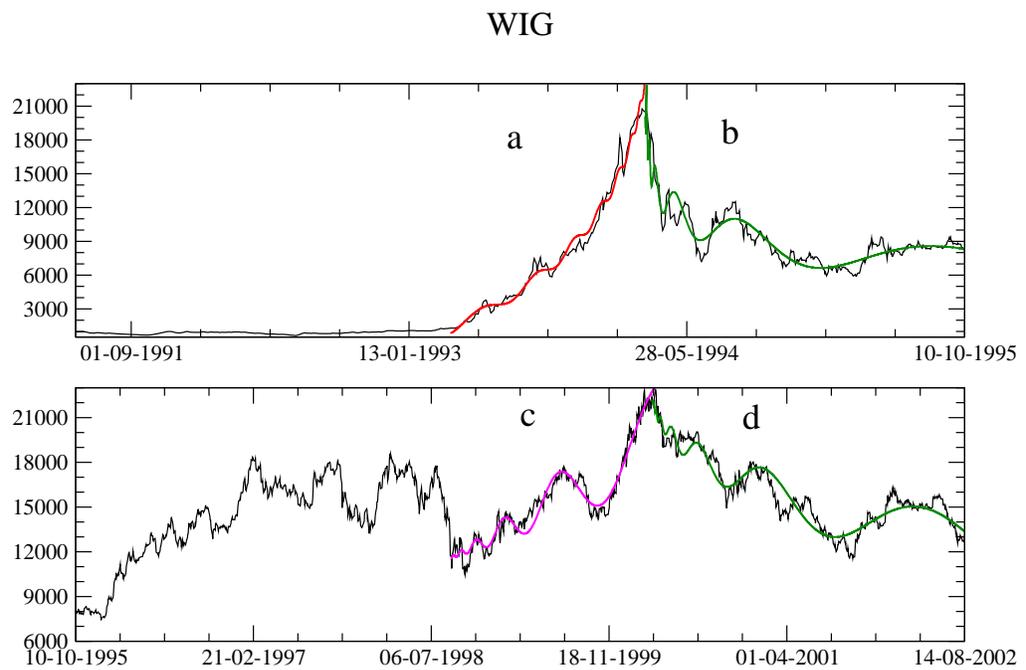}
 \end{center}
 \caption{ 
   Two endogenous crashes observed on Polish stock market index WIG.
   The symbol {\bf a} marks a bubble, the symbol {\bf c} marks an 
   inverted bubble, and {\bf b} and {\bf d} are for anti-bubbles.
	 }
\end{figure}

\begin{figure}
\begin{center}
 \includegraphics[angle=0 , height=16 cm ]{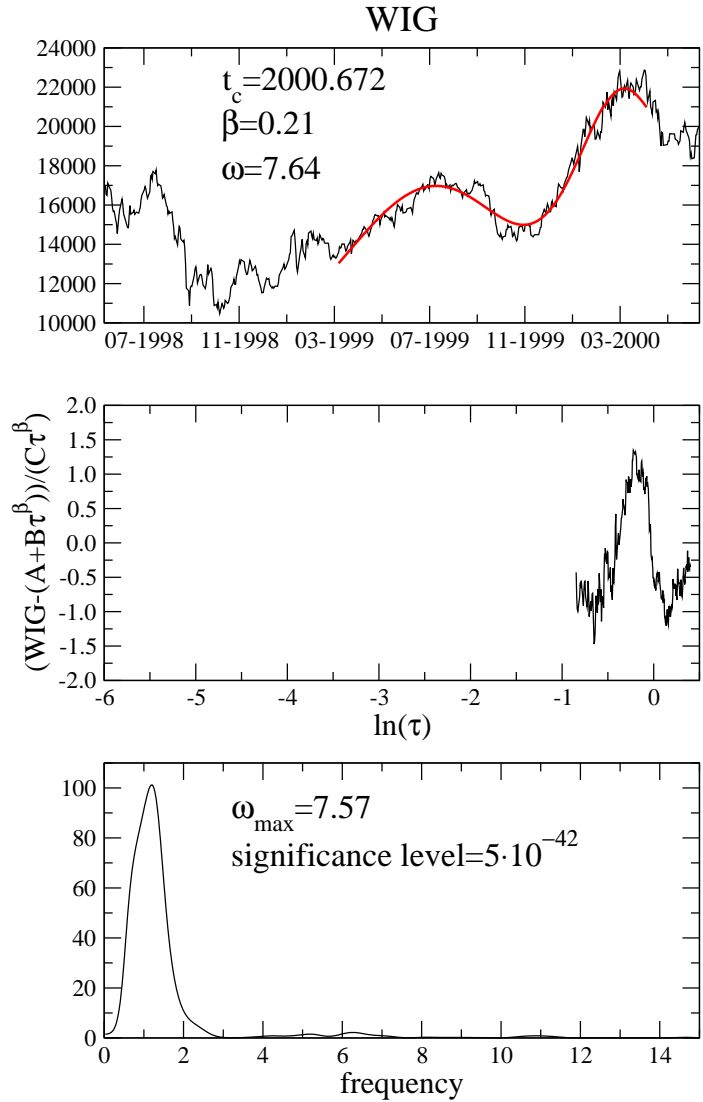}
 \end{center}
 \caption{ 
   Log-periodicity fit to WIG data with crash 4 months after April 2000:
   $\beta=0.21,\ \omega=7.64,\ t_c=2000.672 $.
   The middle panel shows the data transformed to the cosine function with 
   aid of equations (7) and (8). The bottom panel is the Lomb periodogram.
	 }
\end{figure}

\begin{figure}
\begin{center}
 \includegraphics[angle=0 , height=16 cm ]{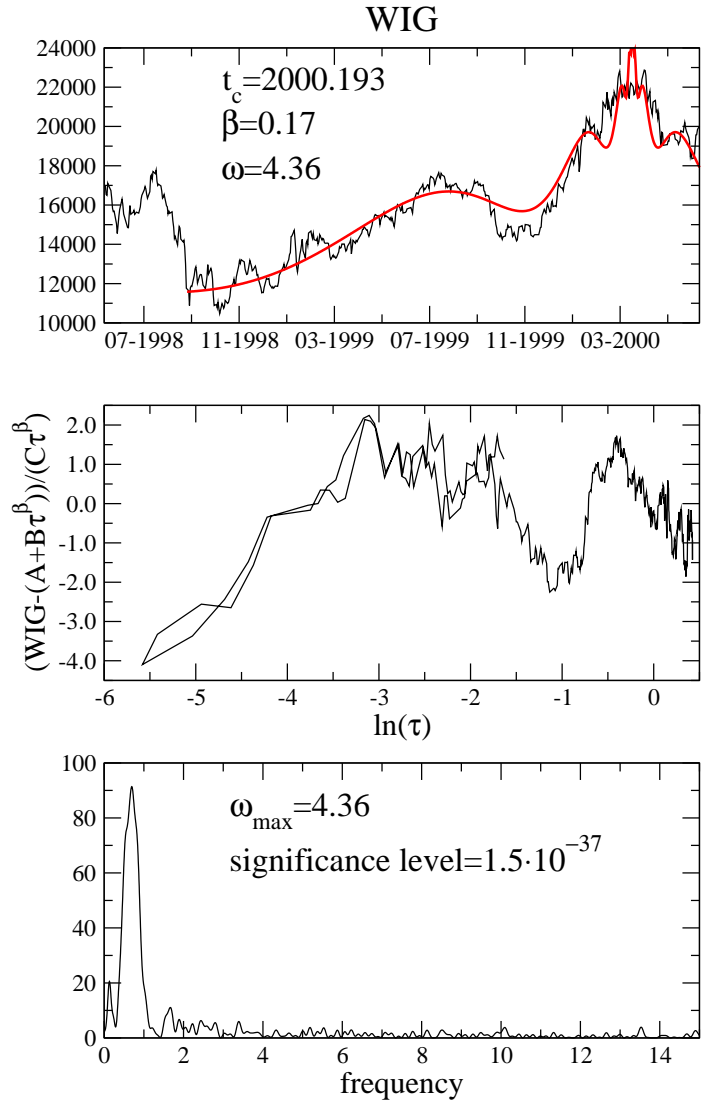}
\end{center}
 \caption{ 
	 A log-periodicity fit to WIG with critical time in April 2000:
         $\beta=0.17,\ \omega=4.36,\ t_c=2000.193$.
         The middle panel shows the data transformed to the cosine function with 
         aid of equations (7) and (8). The bottom panel is the Lomb periodogram.
	 }
\end{figure}

\begin{figure}
\begin{center}
 \includegraphics[angle=0 , height=16 cm ]{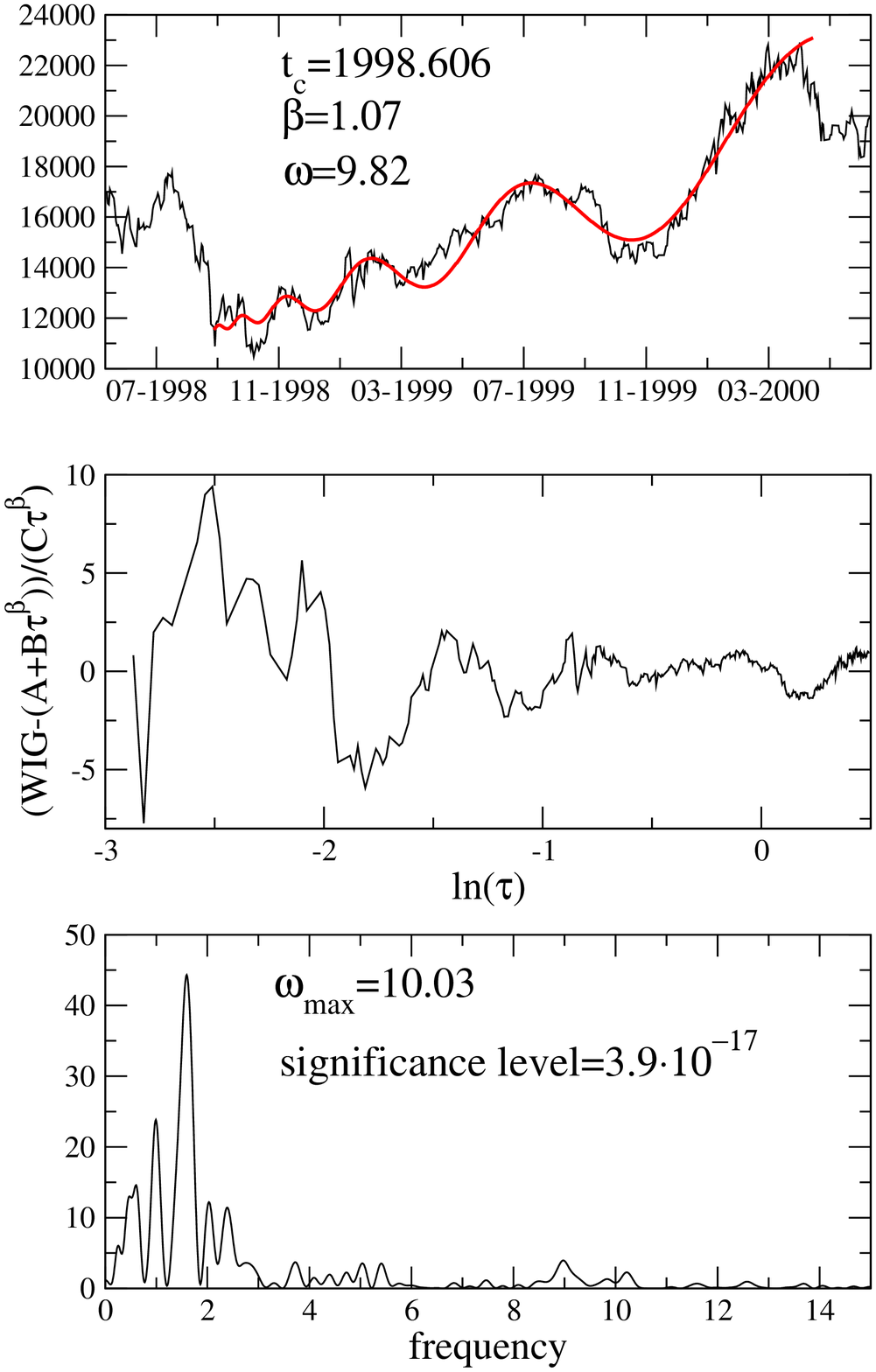}
\end{center}
 \caption{ 
   Inverted bubble for the Polish stock index WIG:
   $\beta=1.07,\ \omega=9.82,\ t_c=1998.606 $.
   Notice, that the fit expands over 4 periods of the cosine 
   function.
   The middle panel shows the data transformed to the cosine function with 
   aid of equations (7) and (8). The bottom panel is the Lomb periodogram.
	 }
\end{figure}


\begin{thebibliography}{}
 \bibitem{First} Sornette, D., Johansen, A., Bouchaud J.P., J.Phys.I France 6, 167-175 (1996)"' sStock market crashes, Precursors and Replicas"'
  \bibitem{Large} Sornette, D., Johansen, A., Physica A 245, 411, (1997) , "Large financial crashes"
  \bibitem{Nasdaq2000} Johansen, A., Sornette, D., EPJ B 17, 319-328, (2000), "The Nasdaq crash of April 2000: Yet another example of log-periodicity in a speculative bubble ending in a crash"
  \bibitem{anti-bubbles} Johansen, A., Sornette, D., Int.J.Mod.Phys.C 10(4), 563-575, (1999), "Financial 'Anti-Bubbles': Log-Periodicity in Gold and Nikkei collapses"
  \bibitem{Bubbles} Johansen, A., Sornette, D., Int.J.Theor.and App. Finance 4 (6), 853-920 (2001), "Bubbles and anti-bubbles in Latin-American, Asian and Western stock markets: An empirical study"
  \bibitem{Exogenous} Johansen, A., Sornette, D., arXiv:cond-mat/0210509, "Endogenous versus Exogenous Crashes in Financial Markets"
  
  \bibitem{predicting} Johansen, A., Sornette, D., O. Ledoit, Journal of Risk, vol. 1, number 4, 5-32 (1999), "Predicting Financial Crashes Using Discrete Scale Invariance"
  \bibitem{comment} Johansen, A., arXiv:cond-mat/0205249, (2002),"Comment on: Are financial crashes predictable?"
  
  \bibitem{outliers} Johansen, A., Sornette, D., EPJ B 1, 141, (1998), "Stock market crashes are outliers"
  
  \bibitem{critical} Johansen, A., Sornette, D., arXiv:cond-mat/9901035, "Critical Crashes"
  
  
  \bibitem{NR} Press, W., Flannery, B., Teukolsky, S., Vatterling, W., "Numerical Recipes in C: the art of scientific computing" , 2nd ed., Cambridge, Cambridge Univ. Press, (1992)
  
  
  
  
  \bibitem{BullishAntiBubbles} Zhou W.-X., Sornette, D., arXiv:cond-mat/0212010, "Evidence of a Worlwide Stock Market Log-Periodic Anti-Bubble Since Mid-2000"

\end{thebibliography}
\end{document}